\documentclass{article}
\usepackage{ijcai22}
\usepackage{xcolor}
\usepackage{times}
\usepackage{soul}
\usepackage{url}
\usepackage[hidelinks]{hyperref}
\usepackage[utf8]{inputenc}
\usepackage{caption}
\usepackage{graphicx}
\usepackage{amsmath}
\usepackage{amsthm}
\usepackage{booktabs}
\usepackage{algorithm}
\usepackage{algorithmic}
\usepackage{amssymb,amsmath,amsthm,dsfont,bm,mathtools,bbm}
\usepackage{pgfplots}
\usepackage{verbatim}
\usepackage[font=small]{caption}
\tikzstyle{basicNode} = [circle, draw=black, fill=white!30]
\tikzstyle{arrow} = [thick,->,>=stealth]
\tikzstyle{transNode} = [circle, draw=none, fill=white!30]
\usepackage{thmtools} 
\usepackage{thm-restate}
\usepackage{tabularx}
\usepackage{multirow}

\usepackage{algorithm}
\usepackage{algorithmic}
\usepackage{cleveref}
\usepackage{xcolor}
\usepackage{booktabs}
\usepackage{subfigure}

\usepackage{color,graphicx}
\usepackage{hyperref,xcolor}
\usepackage[inline]{enumitem}
\usepackage[textsize=tiny, textwidth=1.05in, colorinlistoftodos]{todonotes}

\newcommand{\ALG}[1]{ADVISER}
\title{Deploying ADVISER: Impact and Lessons from Using Artificial Intelligence for Child Vaccination Uptake in Nigeria}

\author{
Opadele Kehinde$^1$\and
Ruth Abdul$^1$\and
Bose Afolabi$^1$\and
Parminder Vir$^1$\and\\
Corinne Namblard$^1$\and
Ayan Mukhopadhyay$^1$\and
Abiodun Adereni$^1$\footnote{Recommended Citation: Kehinde Opadele, Abdul Ruth, Afolabi Bose, Vir Parminder, Namblard Corinne, Mukhopadhyay Ayan, and Adereni Abiodun. ``Deploying ADVISER: Impact and Lessons from Using Artificial Intelligence for Child Vaccination Uptake in Nigeria'', \textit{In Proceedings of AAAI Conference on Artificial Intelligence}, 2024.\\
Corresponding address: ayan.mukhopadhyay@vanderbilt.edu and biodun@helpmum.org}
\affiliations
$^1$HelpMum, Nigeria\\
$^2$Vanderbilt University, USA\\
}

\begin{document}
\maketitle
\begin{abstract}
More than 5 million children under five years die from largely preventable or treatable medical conditions every year, with an overwhelmingly large proportion of deaths occurring in underdeveloped countries with low vaccination uptake. One of the United Nations' sustainable development goals (SDG 3) aims to end preventable deaths of newborns and children under five years of age. We focus on Nigeria, where the rate of infant mortality is appalling. In particular, low vaccination uptake in Nigeria is a major driver of more than 2,000 daily deaths of children under the age of five years. In this paper, we describe our collaboration with government partners in Nigeria to deploy ADVISER: AI-Driven Vaccination Intervention Optimiser. The framework, based on an integer linear program that seeks to maximize the cumulative probability of successful vaccination, is the first successful deployment of an AI-enabled toolchain for optimizing the allocation of health interventions in Nigeria. In this paper, we provide a background of the ADVISER framework and present results, lessons, and success stories of deploying ADVISER to more than 13,000 families in the state of Oyo, Nigeria. 
\end{abstract}

\section{Introduction}

\begin{figure}
    \centering    
    \includegraphics[width=0.9\linewidth]{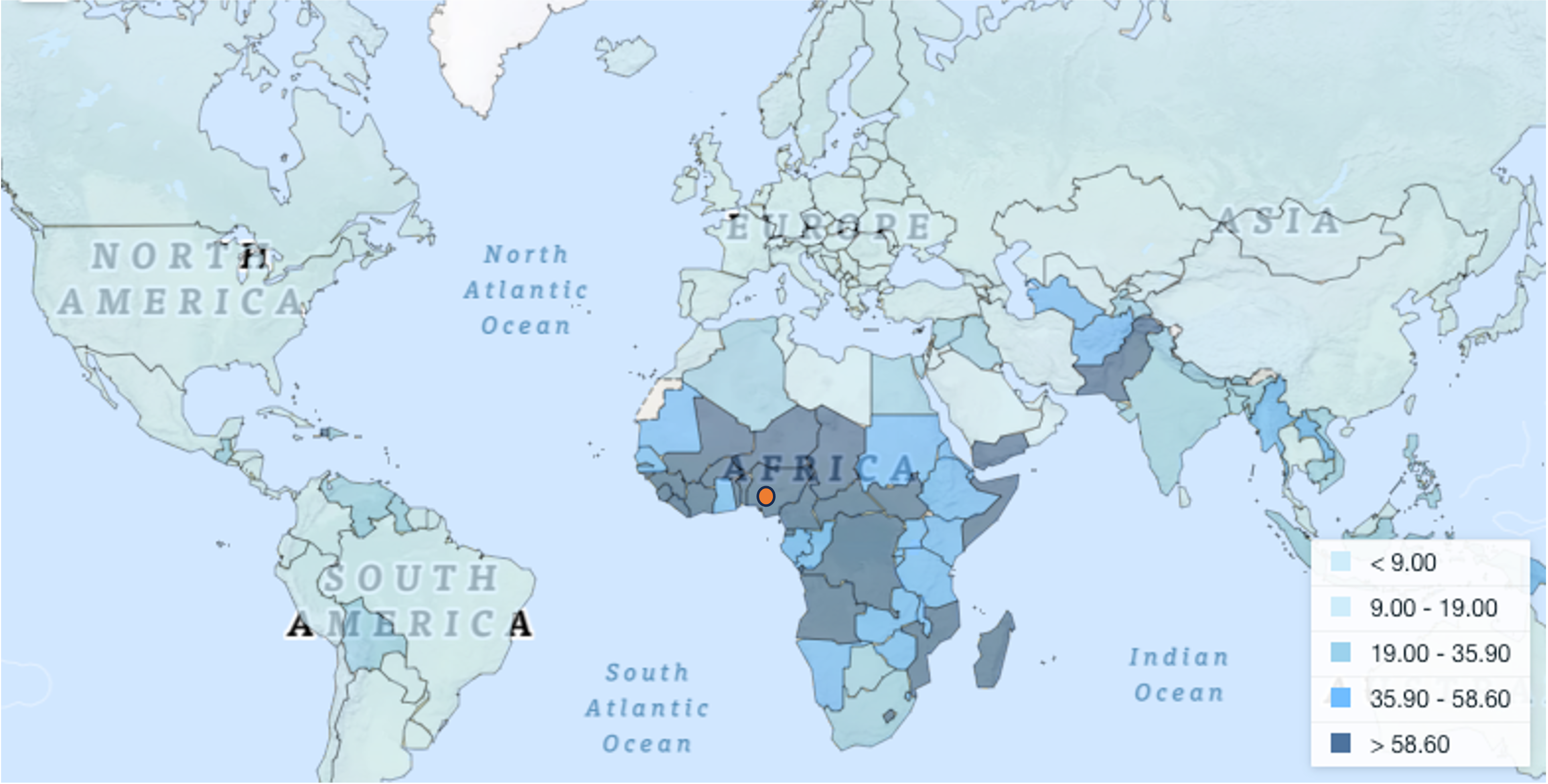}
    \caption{The figure denotes under-five mortality rate, defined as the probability per 1,000 that a newborn baby will die before reaching age five. Nigeria, denoted by the orange point in the map, currently has the third highest rate of child mortality in the world. (Source: World Bank, shared under CC BY-4.0 License)}
    \label{fig:world-map}
\end{figure}

Annually, more than five million children die from medical conditions that are largely preventable or treatable. A disproportionate fraction of these deaths occur in the global south, particularly in countries with low vaccination uptake. We show a global map of child mortality in \cref{fig:world-map}. We focus on Nigeria, where infant deaths in the country are also shockingly high, with more than two thousand daily deaths of children under the age of five years~\cite{okwuwa2020infant}. There are several factors that cause high infant mortality in Nigeria. First, the immunization rate among infants is extremely low despite major efforts by the government. For example, vaccination is available for free through many community health centers. Second (and partly responsible for the first challenge), families often lack the funds required to travel to the community health centers. Finally, while the government and non-profit agencies strive to raise awareness and design interventions to increase vaccination uptake, scaling such interventions has been challenging in practice. 

Addressing maternal and infant health is one of the core imperatives of the United Nations' sustainable development goals (UN-SDGs)~\cite{un_sdg}. Specifically, the SDGs highlight the immediate need to mobilize resources to completely eradicate preventable infant deaths. While mobilizing resources is important, the availability of limited resources (e.g., healthcare centers, community health workers, etc.) means that optimizing the allocation of such resources is critical. We hypothesize that data-driven optimization can particularly benefit in this context, when used in close collaboration with local partners and domain experts. In this paper, we describe the deployment of an AI-enabled optimization engine called ADVISER, an acronym that stands for \textbf{A}I-\textbf{d}riven \textbf{V}accination \textbf{I}ntervention Optimi\textbf{ser}, in Oyo, Nigeria. The ADVISER framework, proposed by \citeauthor{nair2022adviser}~\shortcite{nair2022adviser}, optimizes the allocation of health interventions  in resource-constrained settings. \citeauthor{nair2022adviser}~\shortcite{nair2022adviser} describe how a combination of heuristic search and learning from historical data can design targeted intervention, directing resources to people who need it the most, thereby improving overall health outcomes.

\begin{figure}
    \centering
    \includegraphics[width=0.7\linewidth]{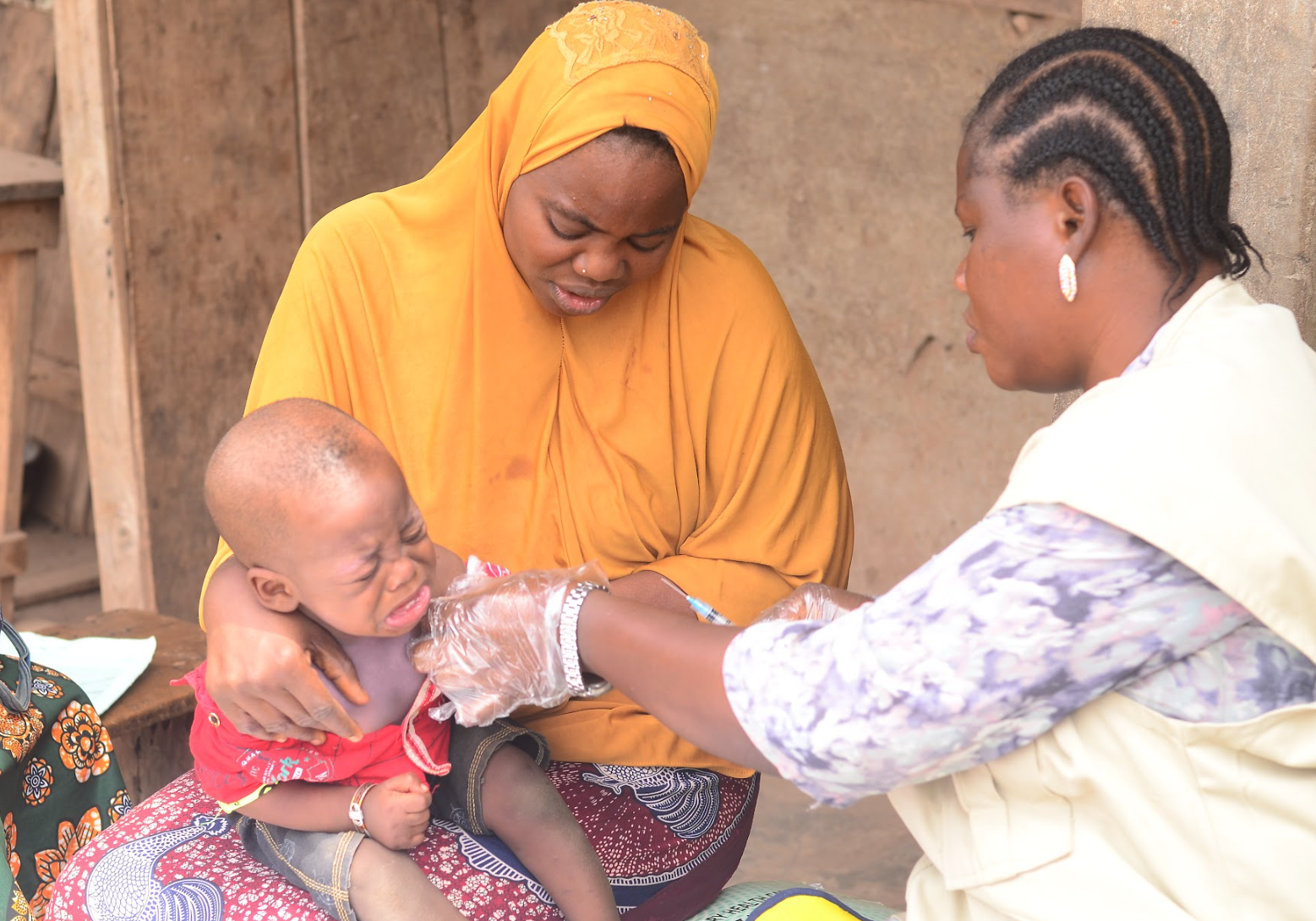}
    \caption{A snapshot from ADVISER's deployment. We envision that the uptake in vaccination through our deployment will potentially pave the way for eliminating infant and child mortality in Nigeria. (Image Courtesy: HelpMum, shared with permission)}
    \label{fig:snapshot}
\end{figure}

While the efficacy of mathematical and data-driven optimization techniques is naturally attractive for resource allocation, deploying such frameworks in practice is particularly challenging. First, data gathering, cleaning, and measuring impact through pre- and post-surveys is often a tedious process. Second, while optimization frameworks can leverage sophisticated computing infrastructure in academic and industrial settings, deployment costs borne by non-profit organizations in the developing world must be low enough to be economically viable. Third, as we show later, hyper-parameters used during algorithm development can often be misleading during deployment, and algorithms must be adapted according to updated settings. Finally, scaling complex socio-technical interventions requires sustained collaborations with local partners, community members, and government agencies. 

In this paper, we describe how we deployed the ADVISER framework to over 13,000 families in the state of Oyo, Nigeria, and addressed these challenges over the course of the deployment. The key components of the deployed framework are:

\begin{enumerate}[leftmargin=*]
    \item To the best of our knowledge, our work presents the first successful deployment of an AI-based framework for optimizing health interventions in Nigeria. We show a snapshot from our deployment in \cref{fig:snapshot}. 
    \item We use the \textit{vaccination tracker}, a database and digital tracker developed and deployed by us, to learn predictive models that inform the optimization engine.
    \item We deploy the optimization framework on cloud infrastructure, develop a lean design, and use approximation techniques to reduce the operational cost of the original ADVISER framework presented by \citeauthor{nair2022adviser}~\shortcite{nair2022adviser}.
    \item We describe challenges faced during the deployment and how the we are adapting the model in response. 
    \item We present results and data from post-deployment surveys to understand the efficacy of the deployment. 
\end{enumerate}
\section{Background}\label{sec:background}

\subsection{Context and Problem Setting}

The Global Vaccine Action Plan (GVAP), supported by the 194 Member States of the World Health Assembly in May 2012, was a principled framework that sought to address millions of preventable deaths by making vaccines available in a more accessible and equitable manner. While Nigeria is also committed to this plan, a nationwide survey
of routine immunization coverage conducted in 2016 showed disappointing results--40\% of children did not receive any vaccines from health system despite living close to a health facility.\footnote{https://www.jhsph.edu/ivac/wp-content/uploads/2018/04/Nigeria-NICS-National-Brief.pdf} The survey also highlighted critical challenges: 42\% of the surveyed families lacked awareness about the importance of immunization, and 23\% mistakenly thought that their child was fully immunized when doses of vaccination were pending. 

We works closely with several local and state governments in Nigeria to address these challenges. For example, it provides clean birth kits to pregnant women in rural areas, trains traditional birth attendants using an e-learning platform and free mobile tablets, and renovates dysfunctional maternity homes to build digital health cafes. In addition, we developed a proprietary vaccination tracking system in 2019 and deployed it to more than 700 health centers in the state of Oyo, Nigeria. The digital database acts as a repository for tracking vaccination---when a mother takes her child for vaccination, a representative logs her (and the child's) details into the tracker using a tablet computer. The tracker is then used to send reminders to the family before upcoming vaccinations through text messages. When the size of the cohort registered in the tracker was relatively low, we also called each family about the importance and the schedule of upcoming vaccinations and the location of the health centers where the vaccine could be delivered.

\subsection{Challenges}

Although reminders and phone calls showed marginal improvement in vaccination uptake, an overwhelming majority of children missed vaccination doses. To understand the challenges faced by the community, we conducted a survey of 1,020 mothers in 2021. Note that while we acknowledge that it is the responsibility of (often both) parents to ensure that children get vaccinated, this responsibility is often spearheaded by mothers in our community; as a result, we chose mothers for the survey, and we also refer to mothers as the beneficiaries of our interventions (in addition to the children who receive vaccination). The survey revealed that while lack of awareness was often a bottleneck, a critical factor that hampered accessibility to vaccinations was the relatively high travel costs to reach public health centers. Almost half the families (46\%) that were included as part of this study earned less than 25 USD per month, and 95\% of the surveyed revealed that a door-to-door delivery of vaccines or travel subsidies would enable them to vaccinate their children. The deployment of ADVISER centers around tackling these challenges. 

\subsection{ADVISER: A Brief Overview}

In response to the survey, we designed four novel interventions for vaccination uptake. The interventions targeted two critical bottlenecks: first, the \underline{lack of awareness} that the community had about the importance of vaccination, and the \underline{the lack of means to access} vaccinations. Specifically, we designed the following interventions:

\begin{enumerate}[leftmargin=*]
    \item Additional Phone Calls: In addition to the text messages and phone calls that we used to make prior to the deployment, we decided to make additional targeted phone calls to mothers who were particularly susceptible to missing vaccination doses for their children. 
    \item Pickup Service: We planned to operate a bus (or a van) service, where we would pick up mothers and children from their residence and drop them off at the nearby health centers. After the vaccination, we would drop them off back to their residence.
    \item Travel Voucher: Since it is infeasible to offer a pickup service to every household, we planned to complement the pickup service with travel vouchers.
    \item Neighborhood Vaccination Drives: Finally, we collaborated with the state government of Oyo for vaccination drive programs, where we would offer a pick up service to a health worker and accompany her on a door-to-door vaccination campaign.
\end{enumerate}

While these interventions promise to increase vaccination uptake, optimizing the interventions presents several challenges. For example, we only have limited workers to make phone calls and operating a pickup and dropoff service is expensive. Also, the state government only has limited number of health workers available for door to door campaigns, as these health workers must attend to their core responsibility of working at local health centers. Finally, the outcomes of the interventions are not guaranteed. For example, providing a travel voucher to a family does not necessarily mean that the voucher would be used for travel to a health center. Similarly, while phone calls can be particularly beneficial for raising awareness, they cannot (by themselves) tackle the problem of accessibility. 

The ADVISER framework optimizes such a set of heterogeneous health interventions by maximizing the cumulative probability of successful outcomes; a successful outcome in our case constitutes an actual vaccination given to a child. We provide a brief overview of the framework here, and refer interested readers to prior work by \citeauthor{nair2022adviser}~\shortcite{nair2022adviser} for specific technical details.\\

\noindent \textbf{Model} The ADVISER framework is based on an ILP that maximizes the cumulative probability of successful vaccinations. First, a data-driven model is learned to estimate the success of each intervention (e.g., a phone call) for each beneficiary, i.e., a mother. We describe this estimation process below. Given the estimates, an ILP is formulated to \textit{match} each individual with \textit{a specific intervention}. The assignment problem naturally imposes integrality constraints. Note that while it is certainly feasible (and perhaps, ideally desirable) to match each beneficiary with more than one intervention, we consider only one match per beneficiary due to budgetary constraints. \\

\begin{table}[!ht]
\begin{tabular}{@{}ll@{}}
\toprule
Age                  & Vaccine                                                                                          \\ \midrule
At Birth             & BCG, HEP B, OPV O                                                                                \\
6 Weeks              & PENT 1, PCV 1, OPV 1                                                                             \\
10 Weeks             & PENT 2, PCV 2, OPV 3                                                                             \\
14 Weeks             & PENT 3, PCV 3, OPV 3, IPT                                                                        \\
6 Months             & VIT A 100 Units                                                                                  \\
9 Months             & \begin{tabular}[c]{@{}l@{}}Measles 1, Yellow Fever, \\ Meningitis\end{tabular}                   \\
12 Months            & VIT A RED 200 Units                                                                              \\
15 Months            & Measles 2                                                                                        \\
18 Months to 5 Years & \begin{tabular}[c]{@{}l@{}}VIT A is administered every 6\\  months in this duration\end{tabular} \\ \bottomrule
\end{tabular}
\caption{List of vaccinations tracked by the ADVISER framework. For each does, we typically follow a time window of 7-10 days centered around the prescribed date of the vaccine. Interventions must be matched within these time windows.}
\label{tab:immunization_list}
\end{table}

\noindent \textbf{Constraints} The ADVISER framework incorporates several domain-specific constraints. For example, each child is eligible for a vaccine for a fixed time period given past vaccinations. We show the vaccinations tracked by the ADVISER system in \cref{tab:immunization_list}. A mother can only be matched to an intervention when her child is eligible for vaccination.\\

\begin{figure*}
    \centering
    \includegraphics[width=0.7\textwidth]{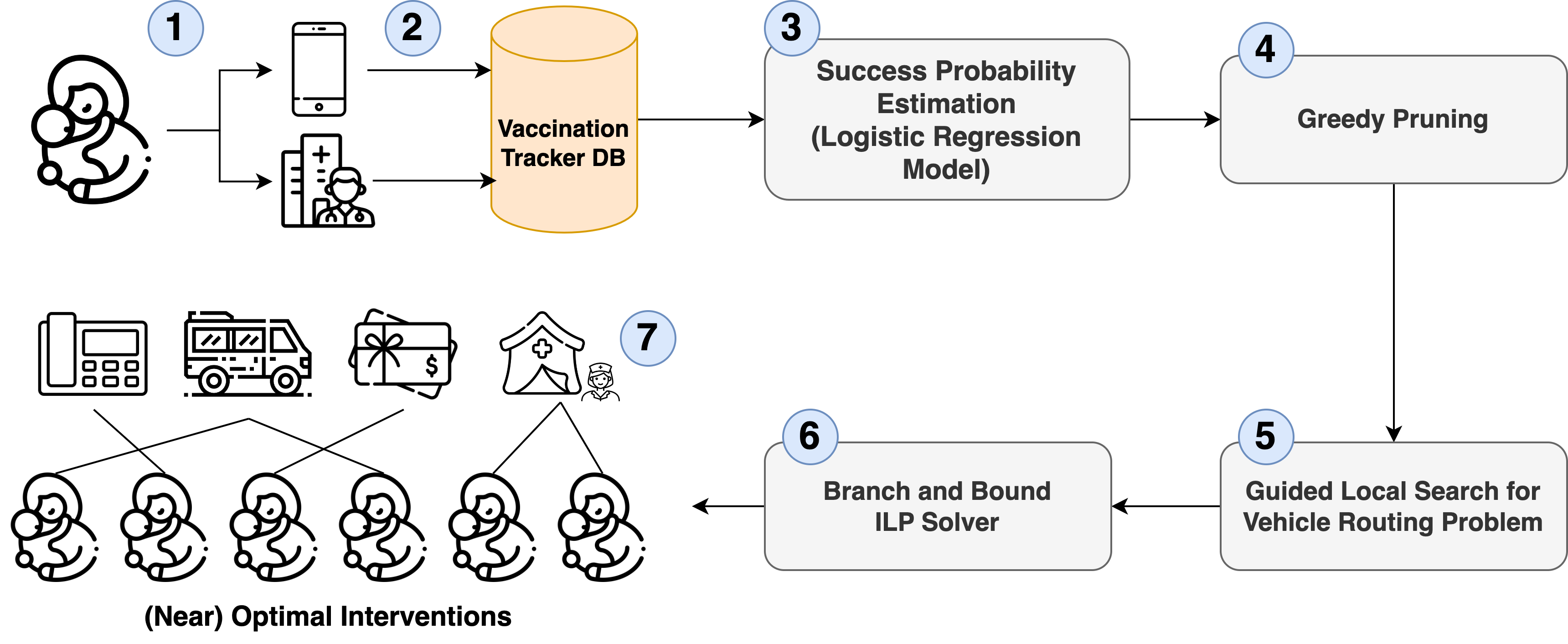}
    \caption{Different stages in the ADVISER framework. 1) We collect data from mothers either through a cell phone application or in-person at health centers and hospitals. 2) The data is cleaned and stored in a database server. 3) We use historical data and surveys to estimate the success of interventions. 4) The resource allocation problem is formulated as an integer linear program, whose solution space is greedily pruned. 5) We do guided local search for generating promising vehicle routes. 6) A significantly smaller ILP, with a reduced solution space, is solved by branch and bound. 7) The matches are then deployed in the field.}
    \label{fig:adviser}
\end{figure*}

Additionally, through discussions with the local health centers and surveys, we found two critical requirements for the interventions: first, the health centers often have long waiting times, and second, individual mothers can have specific times of the day when they are available to take their children for vaccination. Note that mothers must bear this responsibility in addition to employment and additional domestic responsibilities in our community. There are also constraints on the total budget and the availability of resources. We only have four vehicles for the pickup service and we operate under a limited monetary budget. We represent these restrictions as constraints in the ILP.\\

\noindent \textbf{Estimating the Probability of Success} To maximize the probability of successful vaccinations, it is imperative to estimate how likely will an intervention result in a successful outcome. Crucially, the effect of specific interventions can be heterogeneous, e.g., a travel voucher might benefit an individual who knows about the importance of vaccination but cannot afford to access health centers, but might not benefit someone who is not aware of its importance. To estimate these heterogeneous effects, the ADVISER framework represents each beneficiary with a set of attributes or features, such as the mean household income, age of the mother, number of children in the household, and prior vaccination history. Then, the framework uses a logistic regression model to estimate the probability of success for each intervention conditional on the features. Note that this approach presents a \textit{cold-start problem}, as some of the interventions were designed as part of the ADVISER framework itself and we lacked historical data for these interventions (we only had prior data from phone calls). We used a community survey of the 1,020 mothers to estimate these probabilities.\\

\noindent \textbf{Computational Challenges} Solving the ILP in the ADVISER framework is intractable in practice~\cite{nair2022adviser}. There are thee main computational bottlenecks: \textbf{1)} The ILP optimizes over a set of feasible vehicle routes (for picking up mothers en-route the health centers). This is formulated as a vehicle routing problem, and the number of possible routes (i.e., sequence of pickups and dropoffs) exceeds 100 billion with 15-seater vehicles in the city of Ibadan, Oyo. \textbf{2)} The ILP consists over $10^9$ decision variables and $10^9$ constraints. \textbf{3)} While our pilot consists of 13,000 beneficiaries, there are over 1.7 million children in the state of Oyo under the age of five years. Therefore, we must use a scalable approach to solve the ILP. \\

\noindent \textbf{Solution Approach} The ADVISER framework has three major steps. While each step is described in detail by \citeauthor{nair2022adviser}~\shortcite{nair2022adviser}, we present an overview here: \textbf{1)} Since it is infeasible to even generate the set of all feasible routes, we use guided local search to generate promising routes offline, in a process that operates independently of the ILP.  Routes are specifically generated to target mothers that ``need'' them. From a high-level, the need for vaccination is computed from the difference of the probability of successful vaccination \textit{given an intervention} and the probability of successful vaccination \textit{given no interventions}---if this difference rises, intuitively, the need for the intervention increases. \textbf{2)} In order to reduce the search space for the ILP, ADVISER leverages greedy pruning. The pruning is based on a simple idea---if a particular intervention (say, intervention \textit{A}) has a success to cost ratio that is significantly higher than the other interventions, then a part of the budget can be greedily allocated to \textit{A}. In our case, this assumption holds true for vaccination drives. While conducting door-to-door vaccination is relatively more expensive than (say) making a phone call, it guarantees successful vaccination. As a result, we perform greedy pruning based on vaccination drives. \textbf{3)} Given the reduced search space, the ADVISER framework solves a significantly smaller ILP using the branch and bound algorithm~\cite{boyd2007branch}. We show a schematic representation of the ADVISER framework in \cref{fig:adviser}.

\section{Deployment}

We now describe the deployment of the Adviser framework. We begin by describing the bottlenecks we faced to port the framework for deployment and the solution approaches taken to address them.\\

\noindent \textbf{Computational Cost} While the approach proposed by \citeauthor{nair2022adviser}~\shortcite{nair2022adviser} assumes that the exact locations of the mothers are used for computing promising routes for the vehicle routing problem, we found that this was prohibitively expensive in practice. The challenge stems from the fact that our staff collects data manually for the vaccination tracker, i.e., we meet with the mothers at health centers or hospitals and get information such as their home address, phone number, and number of children, among others. However, the addresses are often not complete, and manually geo-mapping the address is extremely labor-intensive (currently, the vaccination tracker has over 90,000 registered mothers). We automated the geo-mapping by leveraging Google Maps API~\cite{maps}.

A larger challenge involved tackling the vehicle routing problem. For generating routes using local search, the ADVISER framework expects as input a travel-time matrix which stores the time taken by a vehicle to travel between any two potential locations of relevance, i.e., each beneficiary's residence and health centers. Unfortunately, computing the travel time matrix is also prohibitively expensive, especially for organizations operating under limited budget in the developing world. For example, during our test deployment with data for 2,558 mothers and 100 health centers, required a total of $(2558+100)^2 = 7064964$ distance queries, amounting to more that USD $\$3000$ in API charges merely during testing (we used the Google Maps API~\cite{maps} for the deployment).

In order to tackle this problem, we leveraged spatial and temporal approximation. First, we divided the entire city of Ibadan into a grid of square cells measuring 1km x 1km. Each location of interest, i.e., a beneficiary's residence or a health center, therefore maps to exactly one cell. We then computed a travel time matrix for the grid using expected operating hours, e.g., as we have to drop off mothers (with their children) at health centers before 11 am (so that there is enough time for vaccinations), we computed the travel time matrix for morning hours. Then, during the local search, this matrix served as a constant time lookup for travel time between any two locations. Crucially, this step involved a one-time cost, i.e., as more mothers get added to the vaccination tracker each day, thereby increasing the number of potential beneficiaries, we can simply map their address to a cell in the grid, without accruing additional costs for calculating travel times.\\

\noindent \textbf{Data} While \citeauthor{nair2022adviser}~\shortcite{nair2022adviser} presented computational results by sampling data from a model trained on 500 mothers, deploying health interventions in practice requires robust estimation of model parameters. For the deployment, we integrated the vaccination tracker database with the ADVISER framework, training the predictive model (that estimates the success of each intervention) with data from over 50,000 mothers (since then, the size of the database has almost doubled). We plan to periodically re-train the models.\\

\noindent \textbf{Software Development} In order to deploy the ADVISER framework in the field, we developed a software around the AI-based optimization engine proposed by \citeauthor{nair2022adviser}~\shortcite{nair2022adviser}. We deployed the AI pipeline on infrastructure
hosted on the Google Cloud Platform (GCP). We wrapped the pipeline as a fully deployable application which takes the following parameters as inputs: 1) time period for generating optimal interventions (e.g., a month); 2) the budget allocated for the time period (e.g., $\$500$); and 3) the model that estimates the success of each intervention. As the staff operating the framework is not familiar with the intricacies of software development and management, the location of the model is pre-set, and staff members only choose the first two parameters. In order to facilitate principled development, deployment, and maintenance, we developed a completely modular software architecture where each major component within the ADVISER framework can be invoked separately, feeding to the next component of the pipeline. 

The first step in the pipeline connects directly to the database of the vaccination tracker to retrieve the set of mothers whose children are eligible for vaccination during the time period chosen for deployment (e.g., the next month). Then, the retrieved data is pre-processed and formatted for the ILP, feeding into the pipeline described earlier. Note that the \texttt{vrp} module accesses a pre-computed travel time matrix instead of computing travel times for each run of the pipeline, saving both time and financial expenses. The ILP is solved by using Gurobi~\cite{gurobi}.\\



\noindent \textbf{Partnerships} The final, and arguably the most critical component for deploying ADVISER involved active collaboration and buy-in from the state government of Oyo, Nigeria. We have had strong collaborations with the state government, having deployed our vaccination tracker as the official immunization tracking application of the state. Our past collaboration paved the way for the deployment of ADVISER. Note that the health officials involved in adopting the ADVISER framework largely lacked background in computer science. As a result, to showcase the promise of the ADVISER framework, we leveraged open-source results and simulations published by \citeauthor{nair2022adviser}~\shortcite{nair2022adviser}. These simulations compared several the state-of-the-art approaches with the ADVISER framework and presented results concerning key health outcomes, such as cumulative number of vaccinations achieved given a fixed budget and the distribution of the interventions. 

For the deployment, we specifically partnered with female nurses and midwives from the Oyo state government. Our discussions with community members and the state government highlighted the importance of this step---often, during the day, mothers might be alone in the house with their children and might not be comfortable with male health workers delivering vaccines. On the other hand, in our community, working with female nurses and midwives has the advantage that they are highly trusted by the local community. The state government, however, asked us ensure that the nurses/midwives are paid a stipend for the field work; we took this cost into account while optimizing the interventions. Naturally, additional costs strain our resources, as we operate as a non-profit organization. Later, we describe how we can make such interventions more financially sustainable.

\section{Results}\label{sec:results}

\begin{figure*}[t]
    \centering
    \subfigure[]{\includegraphics[width=0.23\textwidth, height=3cm]{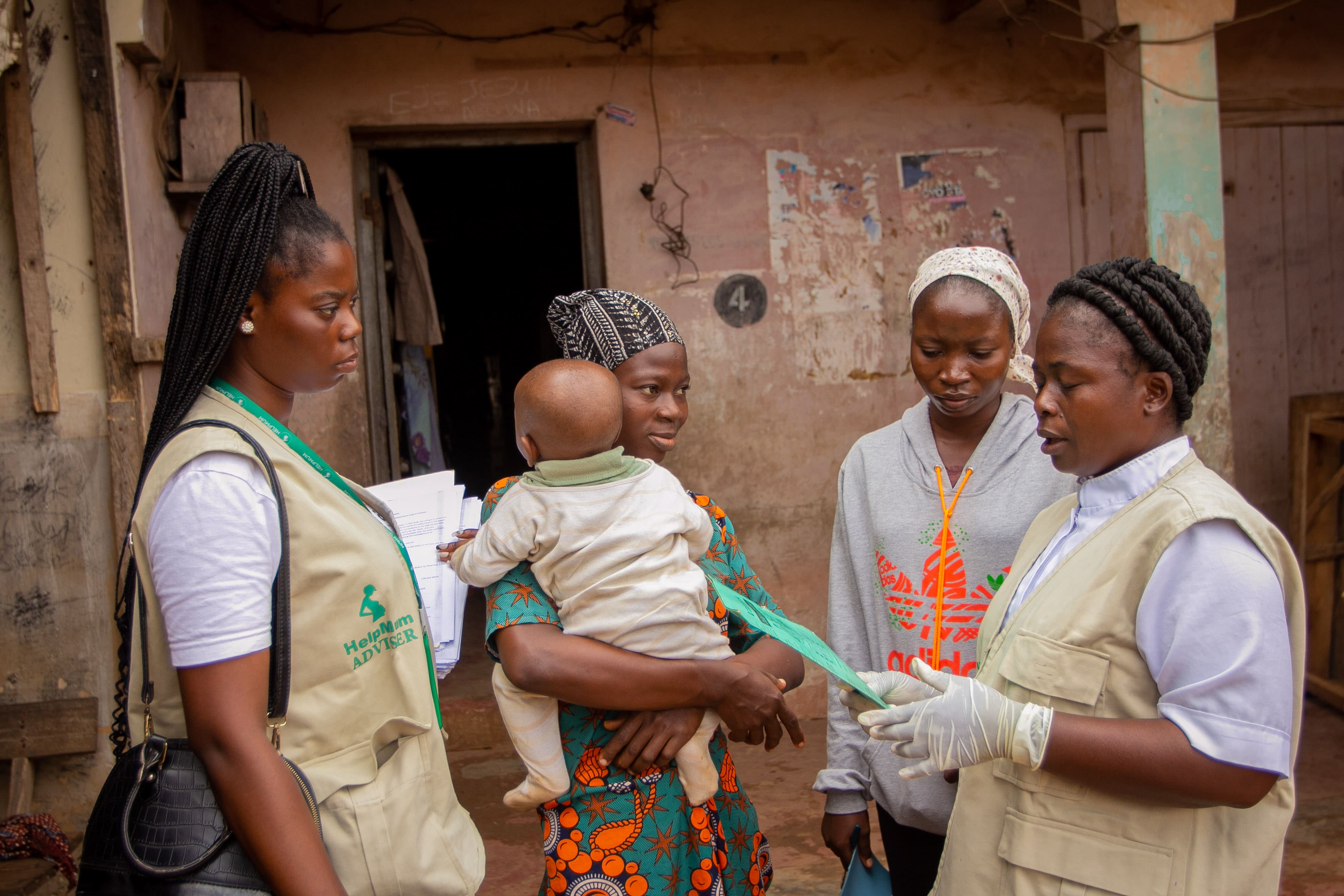}} 
    \subfigure[]{\includegraphics[width=0.23\textwidth, height=3cm]{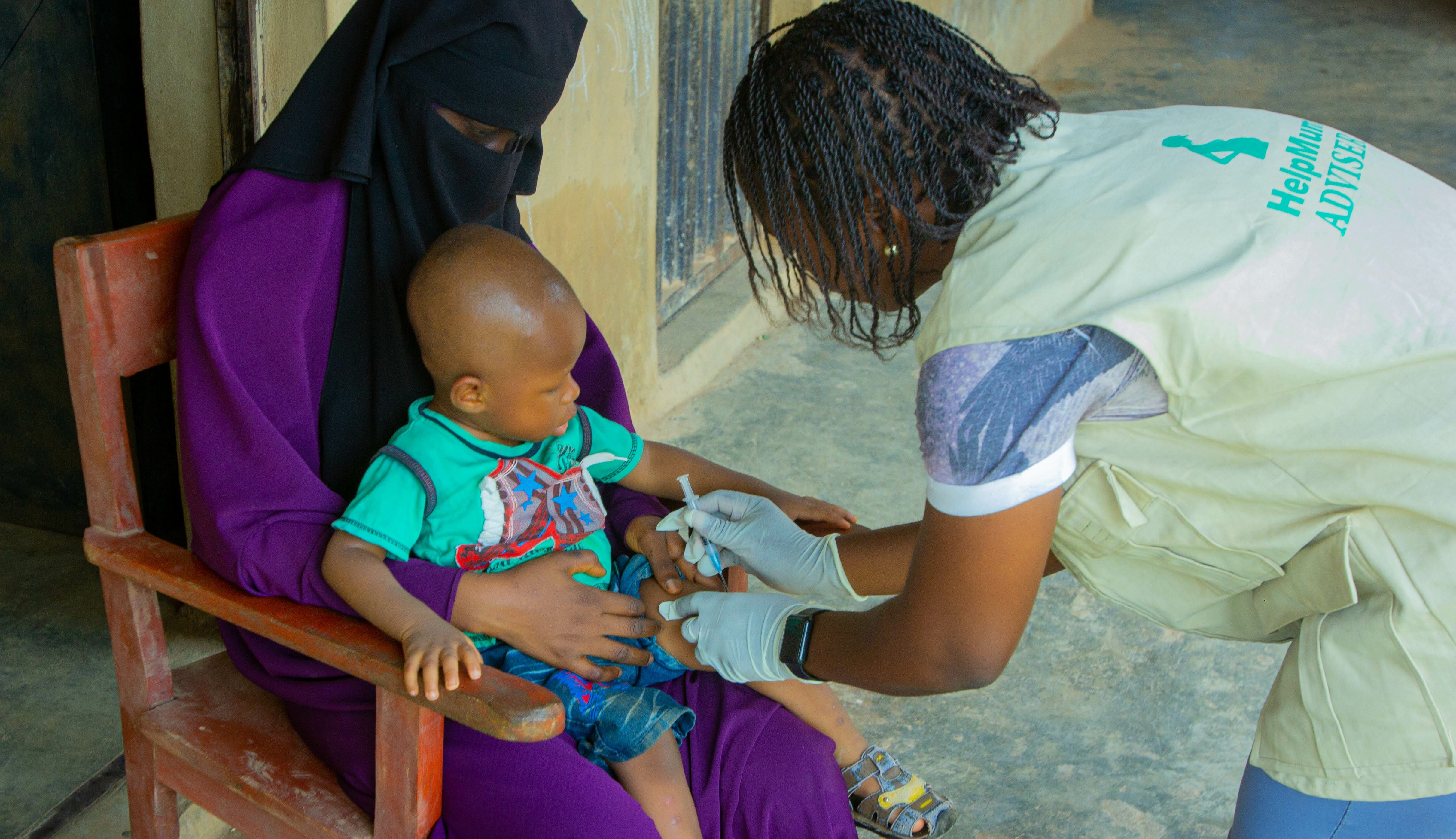}} 
    \subfigure[]{\includegraphics[width=0.23\textwidth, height=3cm]{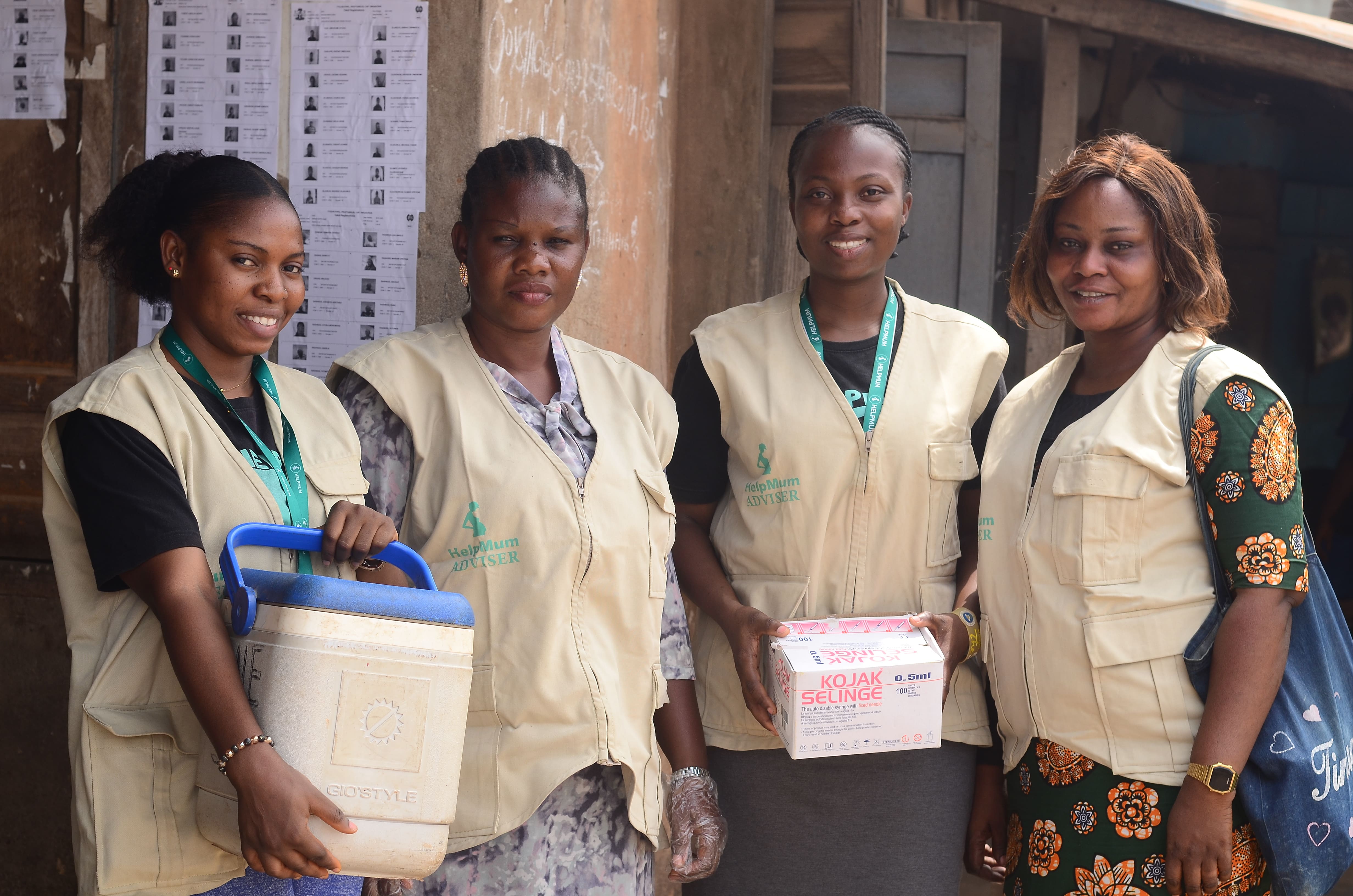}}
    \subfigure[]{\includegraphics[width=0.23\textwidth, height=3cm]{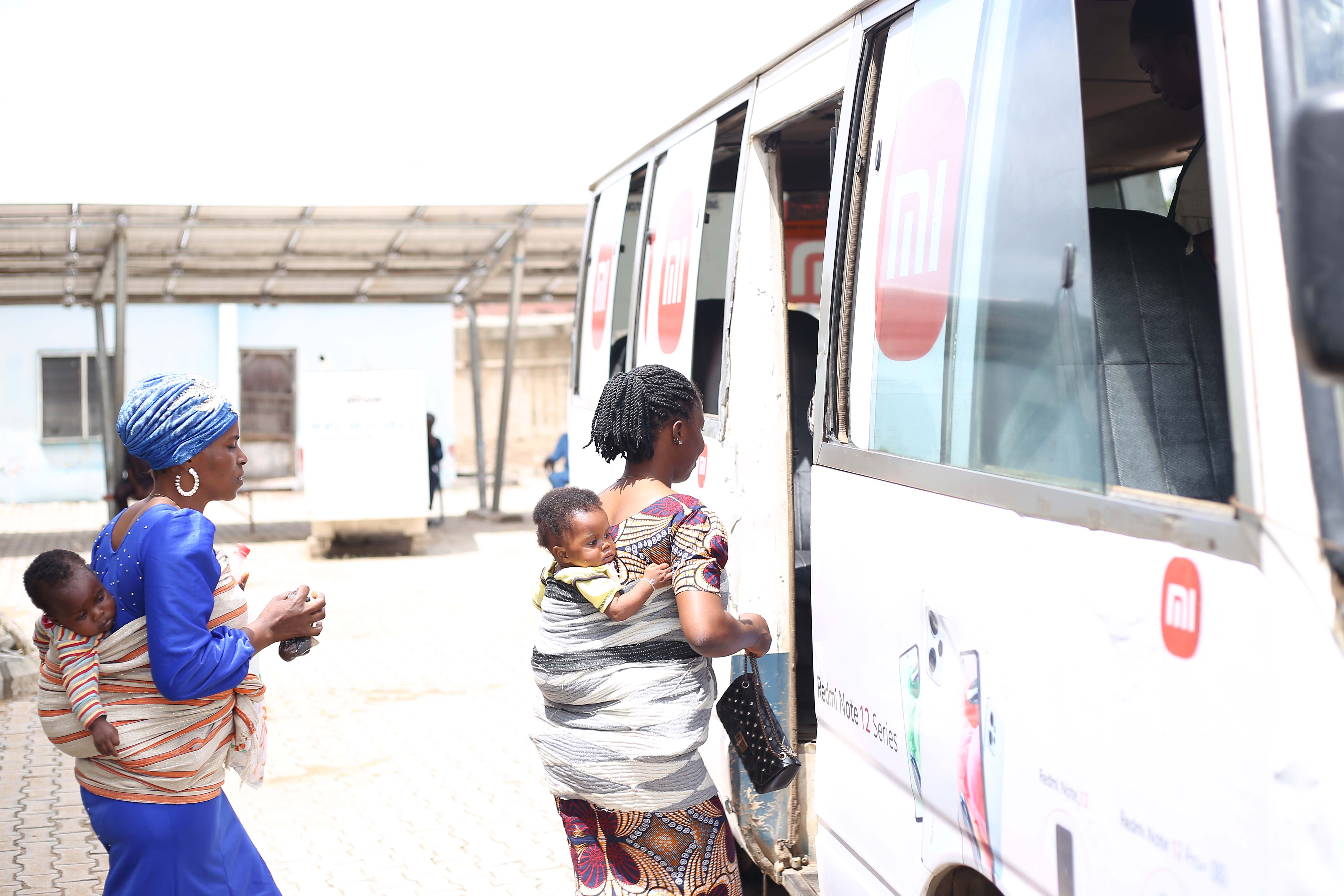}}
    \caption{Images from ADVISER's deployment (all images have been shared with permission) (a) Our staff collecting information to be updated to the Vaccination Tracker database. This data serves as the fundamental building block of the ADVISER framework. (b) A nurse vaccinating an infant during a vaccination drive. (c) Nurses prior to a vaccination drive. The cold storage box (kept in front of the nurses) had limited efficacy during the deployment. (d) Two mothers boarding the pickup service bus with their children. (Images Courtesy: HelpMum, shared with permission)}
    \label{fig:deployment_images}
\end{figure*}

\noindent \textbf{Deployment Setup} We deployed the ADVISER framework in 13 local governments in the state of Oyo, primarily centered in and around Ibadan, the largest city in Nigeria by geographical area. Our deployment was conducted between January 2023--(mid) June 2023. During this period, our interventions were distributed among 13,783 nursing mothers. The local governments were we deployed ADVISER are: Egbeda, Oluyole, Ibadan South West, Ibadan South East, Ona-Ara, Ibadan North East, Ido, Ibadan North West, Lagelu, Ibadan North, Akinyele, Kosofe, Igbogbo/Bayeku, and Ado-Odo/Ota. Out of these, the last three locations are on the outskirts of Ibadan. The thirteen local governments spanned a total of 99 community health centers. Our results are based on deployment in 13 local governments in Nigeria that span a total of 99 community health centers. We show glimpses from the deployment in \cref{fig:deployment_images}.\\

\noindent \textbf{Baseline and Efficacy} To compare the efficacy of the ADVISER framework, we first deployed our pre-ADVISER ``business-as-usual'' approach as baseline, i.e., making phone calls to remind mothers about upcoming vaccinations for their children. We acknowledge that a randomized controlled trial can ideally provide an improved comparison, such a trial is infeasible in the field with nurses/midwives and multiple local governments. We conducted the baseline between August 2022 to December 2022 to all mothers registered in the same 99 community health centers chosen for ADVISER's deployment. We show the cumulative successful vaccination outcomes in \cref{tab:efficacy}. As we see, The ADVISER framework significantly outperformed the baseline during our deployment. Based on the results, our framework has already been approved for deployment in the state of Osun, Nigeria.\\

\noindent \textbf{Post-Deployment Survey} To further analyze the efficacy of the deployment (and understand potential ways of improving the health interventions), we conducted a post-intervention survey with every mother. The survey involved a key question of interest---\textit{was the specific intervention you received critical in getting your child immunized?} The beneficiaries were given five choices, namely \textit{strongly agree}, \textit{agree}, \textit{somewhat agree}, and \textit{disagree}. Note that this survey is critical to a field deployment of targeted interventions; we seek not only to ensure that the beneficiaries get their children vaccinated, but we want to ensure that the interventions are distributed according to the needs of the beneficiaries. For example, if a mother receives a travel voucher but does not need it, a component of the budget is essentially wasted. The feedback gathered through the survey can help in re-tuning the model parameters, thereby improving the distribution of the interventions. We list broad results from the survey below:
\begin{enumerate}[leftmargin=*]
    \item Vaccine Drives: overall, \textbf{93.63\% of the mothers agreed} that the intervention was critical to their children's vaccination. We analyzed that mothers who disagreed with mostly lived close to health centers, suggesting that a in-house campaign was not critical for them.
    \item Travel Vouchers: overall, \textbf{83.92\% of the mothers agreed} that the vouchers were critical. The remaining mothers (mostly) noted that they could have afforded to come to the health center without financial support.
    \item Additional Phone Calls: we observed that \textbf{72.9\% of the mothers agreed} that targeted phone calls were immensely helpful. While this number is (relatively) lower than the other interventions, this behavior is actually expected; recall that every mother registered in the vaccination tracker gets text messages that serve as reminders. The phone calls are \textit{additional}, and as we observe, might not be critical for many beneficiaries. However, since more than 70\% of the mothers found these phone calls to be helpful, we see value in reminders in addition to text messages.
    \item Pickup Service: of the surveyed mothers, \textbf{85.71\% of the mothers agreed} that the pickup service was critical. Similar to the vouchers, the remaining mothers (mostly) noted that they could have afforded to come to the health center without financial support.
\end{enumerate}

\begin{table}[h!]
\centering
\begin{tabular}{@{}lll@{}}
\toprule
Method   & Cumulative Success & Timeline       \\ \midrule
Baseline & 43.6\%             & Aug--Dec 2022  \\
ADVISER  & \textbf{73.9\%}             & Jan--June 2023 \\ \bottomrule
\end{tabular}
\caption{The percentage of beneficiaries whose children received immunization. The ADVISER framework significantly outperformed the baseline during our deployment.}
\label{tab:efficacy}
\end{table}

\begin{table*}[!ht]
\resizebox{\textwidth}{!}{\begin{tabular}{lllll}
\hline
\multicolumn{1}{l|}{Intervention}                          & \multicolumn{1}{l|}{\begin{tabular}[c]{@{}l@{}}Approx.\\ Model \\ Cost\end{tabular}} & \multicolumn{1}{c|}{\begin{tabular}[c]{@{}c@{}}Challenges Faced while \\ Designing ADVISER\end{tabular}}                                                                                                                  & \multicolumn{1}{l|}{\begin{tabular}[c]{@{}l@{}}Approx.\\ Corrected \\ Cost\end{tabular}} & \multicolumn{1}{c}{Lessons from Deployment}                                                                                                                                                                                                                                                                                                                            \\ \hline
\begin{tabular}[c]{@{}l@{}}Phone \\ Calls\end{tabular}     & \$0.4                                                                                & \begin{tabular}[c]{@{}l@{}}We anticipated this to be easiest \\ intervention to administer, based \\ on our prior experience with using\\ phone calls for reminders.\end{tabular}                                         & \$1--2                                                                                   & \begin{tabular}[c]{@{}l@{}}1) Often, cell phones are shared within a family, \\ and mothers are not available when called.\\  2) Moreover, often, multiple calls are required\\ for raising awareness about vaccinations, \\ increasing the cost of intervention per beneficiary.\end{tabular}                                                                         \\ \hline
\begin{tabular}[c]{@{}l@{}}Travel \\ Vouchers\end{tabular} & \$1.1                                                                                & \begin{tabular}[c]{@{}l@{}}Travel vouchers do not guarantee\\ success as the vouchers/cash could\\ be used for other trips.\end{tabular}                                                                                  & \$1.1                                                                                    & \begin{tabular}[c]{@{}l@{}}1) We had planned to distribute vouchers for \\ specific carriers and modes, but community\\ discussions highlighted that people preferred\\ different modes (e.g., bus, tricycle, or taxi).\\ 2) Interestingly, we found that our anticipated\\ voucher of \$1.1 per vaccination worked well.\end{tabular}                                 \\ \hline
\begin{tabular}[c]{@{}l@{}}Vaccine\\ Drives\end{tabular}   & \$15                                                                                 & \begin{tabular}[c]{@{}l@{}}A major bottleneck involves buy-in\\ from the state health centers as \\ vaccines must be procured through\\ the state and administered by trained\\ and trusted nurses/midwives.\end{tabular} & \$20                                                                                     & \begin{tabular}[c]{@{}l@{}}1) Our cost of door-to-door vaccination increased\\ due to steep increase in labor (driver) costs, fuel \\ costs, and stipend considerations. \\ 2) Our initial deployment also led to wastage in\\ resources as the cold-box used for door-to-door\\ drives could only maintain the requisite temper-\\ -ature for 2-3 hours.\end{tabular} \\ \hline
\begin{tabular}[c]{@{}l@{}}Pickup \\ Service\end{tabular}  & \$20                                                                                 & \begin{tabular}[c]{@{}l@{}}The major bottleneck we anticipated\\ was the availability of drivers, \\ vehicles, and long waiting times at\\ the health centers.\end{tabular}                                               & \$30                                                                                     & \begin{tabular}[c]{@{}l@{}}1) Our cost for the pickup service increased due\\ to steep increase in labor (driver) and fuel costs.\end{tabular}                                                                                                                                                                                                                         \\ \hline
\end{tabular}}
\caption{Design Challenges and Cost Corrections based on deploying ADVISER.}
\label{tab:challenges}
\end{table*}

\section{Related Work}

Combinatorial resource allocation for optimizing health outcomes has been applied to several problems, such as streamlining patient admissions in hospitals~\cite{hulshof2013tactical}, managing patients among hospitals in case demand surge (e.g., in case of a pandemic)~\cite{parker2020optimal},  delivering medical and social services to patients in their homes~\cite{aiane2015new}, and data-driven probabilistic frameworks for clinical decision-making~\cite{tsoukalas2015data}. However, the problem we tackle is significantly larger (both in terms of the complexity of the decision space as well a the population under consideration). Recently, AI-based approaches have been used for optimizing health interventions in resource constrained geographies. Two relevant case studies come from India and Kenya; in India, a restless multi-armed bandit model (RMAB) has been used for optimizing phone calls among pregnant mothers~\cite{mate2021field}, and in Kenya, natural language processing has been used for automatically predict the emergency level of pregnant mothers based on the content of their SMS messages~\cite{zhang2023continual}. However, to the best of knowledge, the deployment of ADVISER marks the first real-world deployment of an AI-based method for optimizing the allocation of heterogeneous interventions under uncertainty for vaccination uptake.



\section{Conclusion}

We presented an overview of the ADVISER framework and the results and lessons learned during its deployment to over 13,000 families in Oyo, Nigeria. We believe that this deployment is particularly relevant and important---it is not only the first real-world deployment of an AI-based framework for optimizing vaccination uptake in Nigeria, but it also resulted in a significant increase in vaccination uptake in a region that has one of the highest infant mortality rates in the world. This deployment and the lessons learned would potentially pave the way for improved vaccination outcomes in Nigeria and other parts of Africa. We also hope that this deployment will encourage more collaborative impact-driven projects between non-profit organizations, local governments, and AI professionals.

\section{Acknowledgement}

Authors from HelpMum acknowledge funding from the Patrick J. McGovern Foundation for supporting the deployment of the ADVISER framework. Ayan Mukhopadhyay was supported by the Institute for Software Integrated Systems, Vanderbilt University.

\section{Ethical Statement} We highlight that our approach does not restrict access to vaccinations for any of the beneficiaries. We only optimize additional interventions to ensure that the section of the population that needs interventions the most (due to existing inequities) can get help in receiving vaccination. We are particularly cognizant of potentially harmful effects, and in order to avoid such outcomes, our approach and our deployment are designed in close collaboration with the local governments, community health workers, and our external advisory board.

\newpage

\bibliography{aaai24}

\begin{thebibliography}{}

\bibitem[\protect\citeauthoryear{Aiane \bgroup \em et al.\egroup }{2015}]{aiane2015new}
Daouia Aiane, Adnen El-Amraoui, and Khaled Mesghouni.
\newblock A new optimization approach for a home health care problem.
\newblock In {\em Conference on Industrial Engineering and Systems Management}, 2015.

\bibitem[\protect\citeauthoryear{Boyd and Mattingley}{2007}]{boyd2007branch}
Stephen Boyd and Jacob Mattingley.
\newblock Branch and bound methods.
\newblock {\em Notes for EE364b, Stanford University}, 2006:07, 2007.

\bibitem[\protect\citeauthoryear{{Google}}{2005}]{maps}
{Google}.
\newblock Google maps.
\newblock \url{https://developers.google.com/maps}, 2005.

\bibitem[\protect\citeauthoryear{{Gurobi Optimization, LLC}}{2023}]{gurobi}
{Gurobi Optimization, LLC}.
\newblock {Gurobi Optimizer Reference Manual}, 2023.

\bibitem[\protect\citeauthoryear{Hulshof \bgroup \em et al.\egroup }{2013}]{hulshof2013tactical}
Peter~JH Hulshof, Richard~J Boucherie, Erwin~W Hans, and Johann~L Hurink.
\newblock Tactical resource allocation and elective patient admission planning in care processes.
\newblock {\em Health Care Management Science}, 16(2):152--166, 2013.

\bibitem[\protect\citeauthoryear{Mate \bgroup \em et al.\egroup }{2021}]{mate2021field}
Aditya Mate, Lovish Madaan, Aparna Taneja, et~al.
\newblock Field study in deploying restless multi-armed bandits: Assisting non-profits in improving maternal and child health.
\newblock {\em arXiv preprint arXiv:2109.08075}, 2021.

\bibitem[\protect\citeauthoryear{Nair \bgroup \em et al.\egroup }{2022}]{nair2022adviser}
Vineet Nair, Kritika Prakash, Michael Wilbur, Aparna Taneja, Corinne Namblard, Oyindamola Adeyemo, Abhishek Dubey, Abiodun Adereni, Milind Tambe, and Ayan Mukhopadhyay.
\newblock Adviser: Ai-driven vaccination intervention optimiser for increasing vaccine uptake in nigeria.
\newblock {\em International Joint Conference on Artificial Intelligence (IJCAI)}, 2022.

\bibitem[\protect\citeauthoryear{Okwuwa and Adejo}{2020}]{okwuwa2020infant}
Charles~Onuora Okwuwa and Simon~M Adejo.
\newblock Infant mortality, access to primary health care and prospects for socio-economic development in bwari area council of niger state, nigeria.
\newblock {\em Journal of International Women's Studies}, 21(1):289--308, 2020.

\bibitem[\protect\citeauthoryear{Parker \bgroup \em et al.\egroup }{2020}]{parker2020optimal}
Felix Parker, Hamilton Sawczuk, Fardin Ganjkhanloo, et~al.
\newblock Optimal resource and demand redistribution for healthcare systems under stress from covid-19.
\newblock {\em arXiv preprint arXiv:2011.03528}, 2020.

\bibitem[\protect\citeauthoryear{Tsoukalas \bgroup \em et al.\egroup }{2015}]{tsoukalas2015data}
Athanasios Tsoukalas, Timothy Albertson, and Ilias Tagkopoulos.
\newblock From data to optimal decision making: a data-driven, probabilistic machine learning approach to decision support for patients with sepsis.
\newblock {\em JMIR medical informatics}, 3(1):e3445, 2015.

\bibitem[\protect\citeauthoryear{{United Nations}}{2020}]{un_sdg}
{United Nations}.
\newblock Sustainable development goals.
\newblock \url{https://www.un.org/sustainabledevelopment/health/}, 2020.

\bibitem[\protect\citeauthoryear{Zhang \bgroup \em et al.\egroup }{2023}]{zhang2023continual}
Wenbo Zhang, Hangzhi Guo, Prerna Ranganathan, Jay Patel, Sathyanath Rajasekharan, Nidhi Danayak, Manan Gupta, and Amulya Yadav.
\newblock A continual pre-training approach to tele-triaging pregnant women in kenya.
\newblock {\em Proceedings of the AAAI Conference on Artificial Intelligence}, 2023.

\end{thebibliography}
\bibliographystyle{named}


\end{document}